\documentclass[10pt,conference]{IEEEtran} 
\usepackage{amsfonts}
\usepackage{times}
\usepackage{graphicx}
\usepackage{latexsym}
\usepackage{dsfont}
\usepackage{amssymb}
\usepackage{amsmath}
\usepackage{cite}
\usepackage{verbatim}

\def\bb0{{\mathbb{0}}}

\def\bb{{\mathbf{b}}}

\def\bff{{\mathbf{f}}}

\def\bn{{\mathbf{n}}}

\def\bp{{\mathbf{p}}}

\def\bw{{\mathbf{w}}}

\def\b0{{\mathbf{0}}}

\def\bH{{\mathbf{H}}}
\def\bI{{\mathbf{I}}}

\def\bbE{{\mathbb{E}}}

\def\cE{\mathcal{E}}

\def\sf0{{\mathsf{0}}}

\usepackage{hyperref} 
\usepackage{multirow} 
\usepackage{bigints}
\usepackage{balance}

\DeclareSymbolFont{yhlargesymbols}{OMX}{yhex}{m}{n} \DeclareMathAccent{\yhwidehat}{\mathord}{yhlargesymbols}{"62}

\newcommand{\argmax}[1]{\underset{#1}{\text{argmax}}}
\newcommand{\argmin}[1]{\underset{#1}{\text{argmin}}}

\begin{document} 

\title{Localization in Digital Twin MIMO Networks: \\  A Case for Massive Fingerprinting}

\author{João Morais and Ahmed Alkhateeb \\ \{joao, alkhateeb\}@asu.edu \\ School of Electrical, Computer and Energy Engineering - Arizona State University, USA}

\maketitle
 
\begin{abstract}
Localization in outdoor wireless systems typically requires transmitting specific reference signals to estimate distance (trilateration methods) or angle (triangulation methods). These cause overhead on communication, need a LoS link to work well, and require multiple base stations, often imposing synchronization or specific hardware requirements. Fingerprinting has none of these drawbacks, but building its database requires high human effort to collect real-world measurements. For a long time, this issue limited the size of databases and thus their performance. This work proposes significantly reducing human effort in building fingerprinting databases by populating them with \textit{digital twin RF maps}. These RF maps are built from ray-tracing simulations on a digital replica of the environment across several frequency bands and beamforming configurations. Online user fingerprints are then matched against this spatial database. The approach was evaluated with practical simulations using realistic propagation models and user measurements. Our experiments show sub-meter localization errors on a NLoS location 95\% of the time using sensible user measurement report sizes. Results highlight the promising potential of the proposed digital twin approach for ubiquitous wide-area 6G localization.
\end{abstract}

\section{Introduction}

Precise localization of mobile terminals brings several societal benefits, enabling use-cases both for verticals and consumers, and permitting network design, management, and optimization \cite{LocationAwareComms}. In wireless networks, localization is traditionally done with triangulation (via angle estimation) or trilateration (via estimation of signal propagation time and hence the distance between transmit-receive points) \cite{SurveyLocMethods}. These classical methods mainly work in line-of-sight (LoS); they require the transmission of high-bandwidth reference signals (and thus have high overhead) from multiple access points (AP), and often, these APs need to be synchronized on the order of nanoseconds \cite{SurveyLocMethods}. New localization methods appeared with the introduction of MIMO antenna systems in 4G/5G and Wi-Fi 4 (802.11n) \cite{SurveyLocMethods}. Multiple antennas enabled multipath exploitation techniques, which can work in NLoS and with a single base station \cite{MultipathExploitation, MultipathExploitationAoA}. However, these techniques require specialized hardware and high temporal and spatial resolution, thus imposing high communication overhead and making them unscalable for transparent massive multi-user tracking \cite{SurveyLocMethods}. A promising all-around method that avoids all previously mentioned drawbacks is fingerprinting \cite{RSSI_to_CSI}. 

\textbf{Fingerprinting}:
Fingerprinting consists of determining the user location by pattern-matching measurements with database entries, provided the database has a location associated with each entry \cite{FoundationFingerprinting}. Building the database with the measurements and their locations is arguably the main inconvenience of this method \cite{RSSI_to_CSI}. Some methods have been suggested to reduce the burden of data collection, such as crowd-sourcing \cite{CrowdSourcing} and sensor-based dead-reckoning/tracking \cite{DeadReckoning, WILL_sensorFingerprinting}. However, they have not seen large adoption, which led to reduced research on outdoor fingerprinting \cite{RSSI_to_CSI}. This confirms a long-standing need for approaches demanding low human effort, as identified in \cite{WILL_sensorFingerprinting}. Several works attempted to save human effort by decreasing database size \cite{CrowdSourcing, RSSI_to_CSI}. However, the fundamental principle behind fingerprinting says larger databases are more likely to perform better \cite{FoundationFingerprinting}. 
Fingerprinting localization needs high bijectiveness between features and positions. This raises the question: How can massive databases be built without massive effort?

\begin{figure*}[t]
    \centering
    \includegraphics[width=0.99\linewidth]{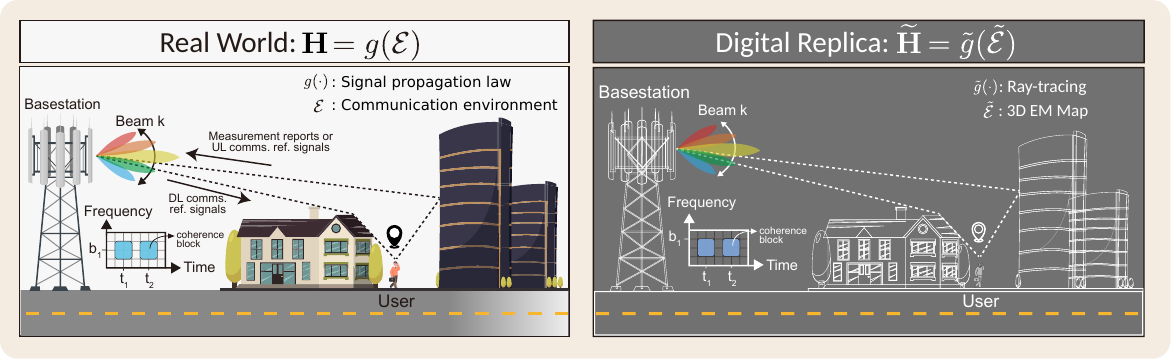}
    \caption{A real-world communication system and its digital replica \cite{ShuaifengDigitalTwin}. For the common case of downlink measurements, the network sends reference signals using certain beams in specific subbands and time instants and the user reports back the received power measurements.}
    \label{fig:model}
\end{figure*}

\textbf{Key Idea}:
This paper proposes digital twin (DT) radiofrequency (RF) maps \cite{DTmagazine_Alkhateeb2023, jiang2024digital} to populate large fingerprinting databases. By leveraging emerging technologies in precise 3D maps and realistic propagation simulations, we create a digital replica of reality and use it to compute RF map fingerprints. This effectively replaces the problem of high human effort required in real-world data collection with the problem of achieving realistic digital replicas. While the first lacks good solutions, the latter is promising and attracting attention. Building database fingerprints with DT RF maps makes localization databases significantly larger and, thus, more performant. Furthermore, near-zero human effort localization allows researchers to think of all possible information that can be included in databases to improve positioning. This work evaluates how RSS measurements from multiple beams and subbands can achieve remarkable positioning accuracy. However, the method can use any simulated information, including multiple base stations, timing/angle/distance data, more channel state information like precoding indicators, and external information like UE inertial sensors or traffic cameras.

\textbf{Contributions}: 
This work proposes DT RF maps as a way to significantly reduce human effort in fingerprinting localization, allowing deployments at scale and new frontiers in positioning accuracy. To the best of our knowledge, this is the first complete database population method and analysis of joint time-frequency-space fingerprinting potential using ray tracing simulations. The method is standard-compliant and has low implementation, ownership, and operation costs. Compared with other localization methods, it does not require LoS, multiple BSs, dedicated hardware, reference signals, or channel estimation \cite{SurveyLocMethods}. It is particularly promising for the sub-6 GHz deployments where bandwidth is scarce, and most users are in NLoS, but its performance improves in higher bands where larger antenna arrays are more available. This work provides a new perspective on digital twin localization.

\textbf{Organization}: Section \ref{sec:Problem} formulates the problem by defining how users measure the received signal strength (RSS) on a given beam and subband and how multiple measurements need to be weighted at the BS to extract a location estimate. Section \ref{sec:Solution} introduces our solution and how 3D maps of the environment form a digital twin of reality where ray tracing simulations compute the propagation paths. The paths are converted into channels and used to build DT RF maps to populate fingerprinting databases. Section \ref{sec:Simulation} details the simulation framework, and Section \ref{sec:Results} evaluates the performance, scalability, and limitations of the proposed solution.

\section{System Model and Problem Formulation} \label{sec:Problem}

Consider a MIMO communication system as represented in Figure \ref{fig:model} where a base station (BS) with $N_t$ antennas communicates with users that are equipped with $N_r$ antennas. Let $\bH \in \mathbb{C}^{N_r \times N_t}$ represent the over-the-air complex channel matrix. Then, the downlink receive signal at the user in a narrowband time-frequency coherence block defined in subband $b$ and time $t$ is given by

\begin{equation} \label{eq:channel_model}
    y[t] = \bw^H \bH_b[t] \bff x[t] + \bw^H \bn,
\end{equation}
with $x$ and $y$ being the transmitted and received symbols in this coherence block, with $x$ agreeing with per-symbol power constraint $\mathbb{E}\left[|x|^2\right] = P_t$, where $P_t$ is the transmit power. The beamforming vector $\bff \in \mathbb{C}^{N_t \times 1}$ follows per-antenna power constraints, i.e., $|f_i| \leq 1$. The vector $\bw \in \mathbb{C}^{N_r \times 1}$ represents the receive combining vector and $\bn \sim \boldsymbol{\mathcal{N}}_{\mathbb{C}^{N_r}}(0,\sigma^2 \bI)$ is the receive noise vector.

\begin{figure*}[t]
    \centering
    \includegraphics[width=1\linewidth]{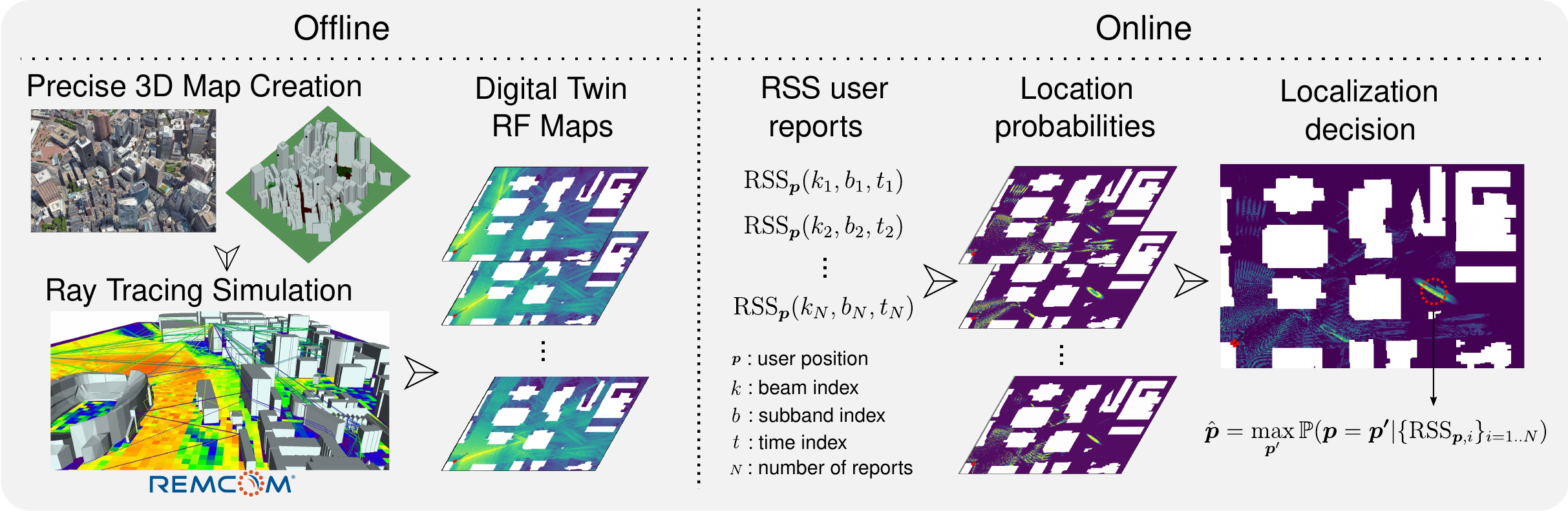}
    \caption{Diagram of end-to-end fingerprinting-based localization system leveraging an offline-built digital twin. The twin is built from ray tracing simulations on all possible user locations on a precise 3D map of the deployment. The twin generates RF Maps for RSS values (or other simulated data) for different beams and subbands, thus replacing the human effort of real-world data collections. These DT RF Maps can be updated or calibrated with real-world information. The maps are used in near real-time with online real-world user measurements to extract location probabilities and make a localization estimate decision. }
    \label{fig:system}
    \vspace*{-.5cm}
\end{figure*}

\textbf{Objective}: We aim to localize the user based on wireless measurements. In particular, the channel matrix $\bH$ in \eqref{eq:channel_model} depends on the user position and the propagation environment. It is, therefore, interesting to investigate the potential of estimating the user position given the channel knowledge \cite{SurveyLocMethods}. Estimating this channel information, however, is challenging in large-scale MIMO systems \cite{ChannelEstimation_Alkhateeb2014}, as it is associated with a large training overhead. Therefore, in this paper, we focus on the user localization problem using more practical wireless measurements, such as received signal strength (RSS) measured across different beams and sub-bands. Next, we define the considered measurement and formulate the user localization problem in terms of these measurements. 

\textbf{RSS Measurement Set}: With the downlink transmission, we assume that the user will measure the RSS across multiple BS transmit beams, subbands, and time blocks and feed these measurements back to the BS. Let $b$ and $t$ be the subband and time indices that localize a time-frequency coherence block and $k$ be the index of the BS beam $\bff_k$ in the codebook $\boldsymbol{\mathcal{F}}$ where the RSS measurement takes place. For ease of exposition, we assume single antenna user terminals, i.e. $N_r = 1$. Therefore, the RSS measured in the coherence block $(b,t)$, a user position $\bp$ and BS beam $k$ can be written as 

\begin{equation} \label{eq:RSS}
    \text{RSS}_\bp(k,b,t) = \left| y_\bp(\bff_k,\bH_b[t]) \right|^2.
\end{equation}
When many of these measurements take place in different beams, subbands, and time slots, the result is a measurement set $\eta_\bp$, which can be formulated as 
\begin{equation} \label{eq:report}
    \eta_\bp(\mathcal{K}, \mathcal{B}, \mathcal{T}) = \left\{ \text{RSS}_\bp(k, b,t) : k \in \mathcal{K}, b \in \mathcal{B}, t \in \mathcal{T} \right\}
\end{equation}
where $\mathcal{K}$, $\mathcal{B}$, and $\mathcal{T}$ are, respectively, the sets of beams, subbands, and time blocks where measurements occurred. Note that the measurements should be on a timescale where the position can be considered constant. Otherwise, the measurements should not be aggregated together to estimate a single position. Note also that larger sets ($\mathcal{K}$, $\mathcal{B}$, or $\mathcal{T}$) hold more information about the user position. Hence, this is expected to lead to higher localization performance. 

\textbf{Reported Beams}: 
Regarding the specific beams reported by the user (the beams in the set $\mathcal{K}$), users typically report the received power in the best-received beams \cite{BeamManag_Giordani2019}.  If the BS codebook size is denoted by $|\boldsymbol{\mathcal{F}}|$, then we can define this set of the $N_K$ highest received power beams as 
\begin{equation} \label{eq:K}
    \mathcal{K} = \underset{|\mathcal{K'}| = N_K}{\argmax{\mathcal{K}' \subset \{1 .. |\boldsymbol{\mathcal{F}}|\}}} \ \sum_{k \in \mathcal{K}'} \text{RSS}_\bp(k,b,t).
\end{equation}

\textbf{Problem Definition}:
With the set of measurements $\eta_\bp$ defined, the use can be localized using $\mathcal{L}^{'} \left(\eta_\bp(\mathcal{K}, \mathcal{B}, \mathcal{T})\right)$, where $\mathcal{L}^{'}$ is a mapping (localization) function that can localize the user given the reported measurements. Our objective in this paper is to learn/optimize this mapping function. If $\mathcal{P}$ represents the position dataset collected for optimizing the mapping/localization function optimization (for example, the fingerprinting database), we can formulate our objective as
\begin{equation} \label{eq:pos_mapping}
    \mathcal{L} =  \argmin{\mathcal{L}'} \ \bbE_{\mathcal{P}} \left| \bp - \mathcal{L}^{'} \left(\eta_\bp(\mathcal{K}, \mathcal{B}, \mathcal{T})\right) \right| ,
\end{equation}
where $\mathcal{L}$ represents the target mapping function (e.g., k-nearest-neighbors) we want to design to make the most accurate estimate for the user position from the sets of measurements $\eta_\bp$. Note that this problem formulation is generally applicable to any MIMO system, including current Wi-Fi and cellular networks. 

\section{Digital Twins for Localization} \label{sec:Solution}
The design of a localization function $\mathcal{L}$ that maps a set of measurements $\eta_\bp$ to a user location requires building an initial domain of measurement data, typically demanding laborious real-world data collections in the area of interest. In contrast, this work proposes to leverage a digital twin of the environment to build \textit{synthetic} RF maps, i.e., a fingerprinting database, thus eliminating most human labor involved. A DT can be created using the 3D maps of the environment, ideally fused with the material properties and the real-time dynamics obtained from various sensing information \cite{DTmagazine_Alkhateeb2023}. Subsequently, we can obtain an approximation of real propagation by performing electromagnetic simulations, such as ray tracing, in this digital replica. This section describes the proposed approach in detail.

\textbf{Digital Replica}: First, we adopt a similar formulation to \cite{ShuaifengDigitalTwin} for the digital twin approximation. Let $\widetilde{\cE}$ be our 3D  model approximation of the real world $\cE$ (including the material characteristics). Likewise, ray tracing is our approximation $\widetilde{g}(\cdot)$ of the propagation laws of nature $g(\cdot)$. Then, we may write the real and digital wireless channels as

\begin{equation} \label{eq:digital_channel}
    \bH = g\left(\cE\right) \leftrightarrow \widetilde{\bH} = \widetilde{g}\left(\widetilde{\cE}\right).
\end{equation}

\textbf{Channels From the Digital Replica}:
The 3D model and ray tracing are exploited to construct the channels in the digital replica. The 3D models, currently static but in the future updated in real-time, are used in ray-tracing to generate channel parameters such as angles of arrival (AoA) and angles of departure (AoD) for each path propagating from the transmitter to the receiver. Then, a geometric channel model can be utilized to construct the channel matrix: The approximation of the channel impulse response $\widetilde{h}_{i,j}(t)$ between a transmit-receive antenna pair in the digital replica can be written as the sum of $L$ multi-path components
\begin{align}\label{eq:ray_tracing}
    \widetilde{h}_{i,j}(t) = \sum_{l=1}^{L}\alpha_l \delta(t-\tau_l) G_{\mathrm{i}}\left(\phi^{\mathrm{AoA}}_l, \theta^{\mathrm{AoA}}_l\right) G_{\mathrm{j}}\left(\phi^{\mathrm{AoD}}_l, \theta^{\mathrm{AoD}}_l\right),
\end{align}
where $\alpha_l$ and $\tau_l$ represent the complex gain and propagation delay of the $l$-th path, and the azimuth and elevation angles of arrival and departure of this path are respectively denoted by $\phi^{\mathrm{AoA}}_l$, $\theta^{\mathrm{AoA}}_l$, $\phi^{\mathrm{AoD}}_l$ and $\theta^{\mathrm{AoD}}_l$. $G_{\mathrm{i}}$ and $G_{\mathrm{j}}$ are the radiation patterns of the receive and transmit antennas.

\textbf{DT RF Maps}:
The channels can be used to populate a DT database according to Equation \eqref{eq:RSS} with the simulated RSSs denoted by $\widetilde{\text{RSS}}_\bp(k,b)$. This database has dimensions $[\mathsf{D}_K, \mathsf{D}_B, \mathsf{D}_P]$, where $\mathsf{D}_K = |\boldsymbol{\mathcal{F}}|$ is the number of BS beams, $\mathsf{D}_B$ is the number of subbands and $\mathsf{D}_P$ is the number of simulated positions in a 3D user grid. This database represents synthetic DT RF maps with simulated RSS values. 

\textbf{Localization Probability:} 
When a user reports the real-world measurement $\text{RSS}_\bp(k,b,t)$, the DT RF maps are used to compute a 3D probability grid of where the user is more likely to be in the position space. This is achieved by computing the overall perceived probability of a user being in position $\bp'$ given the set of measurements $\eta_\bp(\mathcal{K}, \mathcal{B}, \mathcal{T})$, according to
\begin{equation} \label{eq:2d_prob} 
    \mathds{P} \left(\bp = \bp' \ |\ \eta_\bp(\mathcal{K}, \mathcal{B}, \mathcal{T})\right).
\end{equation}

This probability distribution is calculated based on measurements from the DT, and a possible approach is described in Section \ref{sec:Simulation}. 

\begin{figure}[t]
    \centering
    \includegraphics[width=1\linewidth]{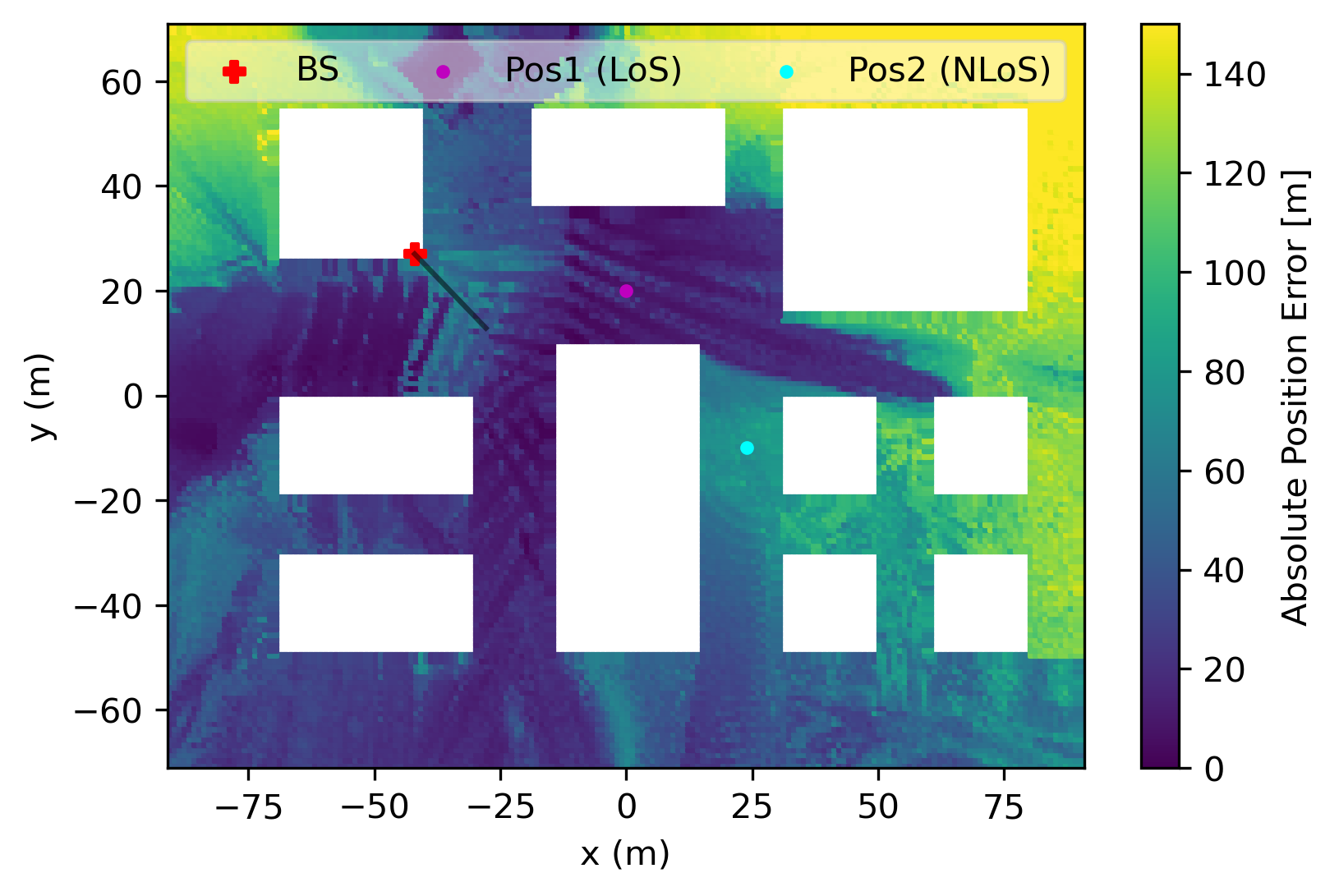}
    \caption{Localization accuracy for every possible user position in a 2D grid at a height of 2 meters (6.5 feet). The simulation was performed for the proof of concept scenario with 6 buildings using the baseline reporting parameters: no. beams $|\mathcal{K}| = 1$, no. subbands $|\mathcal{B}|=1$, and no. of reports in time $|\mathcal{T}|=1$. BS uses a 64-antenna ULA with position and orientation represented. Positions 1 (50m from BS, LoS) and position 2 (80m from BS, NLoS) are used next.}
    \label{fig:2D_acc_grid}
    \vspace*{-.3cm}
\end{figure}

\textbf{Location Estimate}:
Figure \ref{fig:system} summarizes at a high level the localization workflow. To recapitulate, we use equation \eqref{eq:2d_prob} to obtain the location probability of a user being in position $\bp'$ given a set of real-world measurements $\eta_\bp$. Each measurement in the set gives rise to a location probability, which we can compute by assuming a measurement probability distribution composed of real-world measurements and/or digital twin results. Following these steps, we obtain the likelihood of a user being in position $\bp'$. To find the position estimate $\yhwidehat{\bp}$, the position likelihood computation is repeated for all positions in the DT database, to see which leads to a higher probability. Mathematically, the process follows

\begin{equation} \label{eq:pos_estimation}
    \yhwidehat{\bp} = \yhwidehat{\mathcal{P}} \left(\eta_\bp(\mathcal{K}, \mathcal{B}, \mathcal{T})\right) = \underset{\bp'}{\max} \ \mathds{P} \left(\bp = \bp' | \eta_{\bp'}(\mathcal{K}, \mathcal{B}, \mathcal{T})\right),
\end{equation}
where we denote $\yhwidehat{\mathcal{P}}$ as the positioning function found via probability based method. According to the information in the DT database, this localization function attempts to minimize the localization error given by
\begin{equation} \label{eq:pos_error}
    \epsilon = \left| \bp - \yhwidehat{\bp}\right|,
\end{equation}
for a certain set of measurements $\eta_\bp$ (subject to the DT modeling accuracy). Now, we evaluate how larger measurements/fingerprints relate to higher accuracy, given our proposed solution. If this proves to be true, then we can conclude that this method has the potential for providing very accurate positioning while avoiding human effort.

\section{Simulation Setup} \label{sec:Simulation}

After theoretically determining a strategy to obtain the user location based on a set of measurements $\eta_\bp(\mathcal{K}, \mathcal{B}, \mathcal{T})$, we evaluate the location accuracy of the proposed DT based localization approach and how does the measurement set size affect this accuracy \footnote{Code available in \href{https://github.com/jmoraispk/LocalizationDigitalTwins}{https://github.com/jmoraispk/LocalizationDigitalTwins}}. We hypothesize that more independent measurements, along any dimension, lead to better accuracy. The three dimensions considered are beams reported ($|\mathcal{K}|$), reported subbands ($|\mathcal{B}|$), and reports in time ($|\mathcal{T}|$). Conducting this experiment requires building a digital twin database. 

\textbf{Database Creation}: 
We perform ray tracing simulations using Wireless Insite \cite{InSite} on a proof of concept scenario with 6 buildings. The DB grid spans an area of 180 x 140 meters at a height of 2 meters with a resolution of 2 meters. The result is a uniform grid of 91 by 71 positions, 6461 in total. We exclude the positions inside buildings, narrowing the possible user positions down to 4286. The top view of this scenario is present in figure \ref{fig:2D_acc_grid}. The ray tracing simulations use a depth of five reflections with enabled scattering and diffusion but no material penetration. The simulator sends rays from the BS in 5 million directions to detect the power contributions from each path towards the user.

\textbf{Channel Generation}:
After ray tracing the paths, we use DeepMIMO \cite{DeepMIMO} to generate the OFDM channels. These channels have 15 kHz subcarriers (numerology 0 in 5G) that span a bandwidth of 20 MHz. We aggregate the channel response of these subcarriers in subbands of 1 MHz, and repeat for all positions and beamformers/beams we want to evaluate. 

\textbf{RSS Distributions}: 
For an initial analysis, instead of collecting real-world RSS measurements, we resort to sampling this data from random distributions. We assume the measured RSS follows a normal distribution, i.e. $\text{RSS}_\bp(k,b,t) \sim \mathcal{N}\left(\mu_\bp(k,b), \sigma_\bp(k,b)\right)$. Assuming that ray tracing simulations are realistic, we can define $\mu_\bp(k,b) = \widetilde{\text{RSS}}_\bp(k,b)$, where $\widetilde{\text{RSS}}_\bp(k,b)$ is the ray-traced RSS value for position $\bp$ in beam $k$ and band $b$. For the standard deviation, $\sigma_\bp(k,b) = \sigma_{def}$ with $\sigma_{def} = 2 $ dBm. This choice guarantees 99\% of measurements within $\pm 6$ dB of the mean, which agrees with reported variations between 5 and 7 dB for immobile users \cite{RSSI_to_CSI}. 

\begin{figure}[t]
    \centering
    \includegraphics[width=0.985\linewidth]{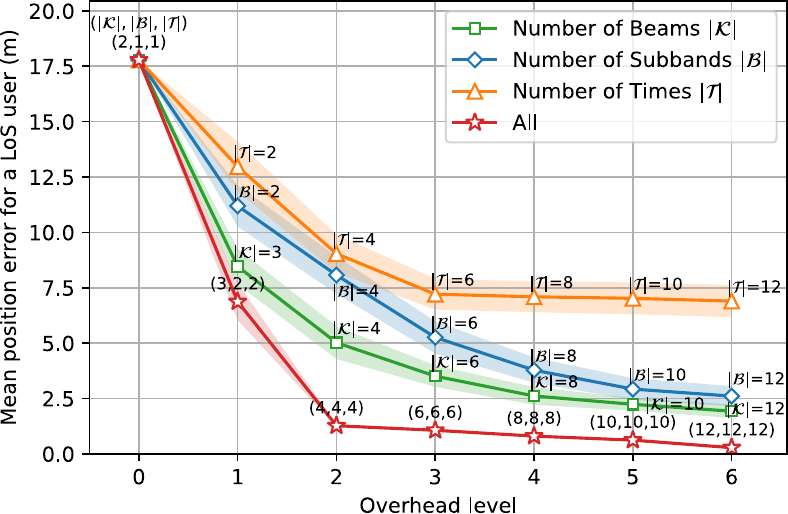}
    \caption{Impact of several reporting parameter combinations on localization error for position 1 (LoS). RSS measurements in more beams ($|\mathcal{K}|$), subbands ($|\mathcal{B}|$) and repeated measurements across time ($|\mathcal{T}|$) individually and jointly improve fingerprinting accuracy.}
    \label{fig:pos1_multi}
    \vspace*{-.5cm}
\end{figure}
\textbf{Position Computation:} To compute the location of each user, the probability likelihood on the right-hand side of equation \eqref{eq:pos_estimation} needs to be defined. The probability likelihood depends on the assumptions on the wireless channel. It is known that the wireless channel is correlated in time, frequency and space. However, this correlation is complex and hard to take into account. For this reason, initial analysis typically considers independent fading realizations in consecutive coherence intervals. This assumption makes our results approximate a lower bound on localization accuracy because the correlated information across measurements is not used. Under the assumption that measurements $\text{RSS}_\bp(k,b,t)$ performed in different beams and different coherence blocks are independent, the location probability likelihood of a set of measurements $\eta_\bp(\mathcal{K}, \mathcal{B}, \mathcal{T})$ can be given by

\begin{equation} \label{eq:2d_prob2} 
    \mathds{P} \left(\bp = \bp' \ |\ \eta_\bp(\mathcal{K}, \mathcal{B}, \mathcal{T})\right) = \hspace{-.5cm} \prod_{\substack{k \in \mathcal{K}, b \in \mathcal{B} , t \in \mathcal{T}}} \hspace{-.5cm} \mathds{P} \left(\bp = \bp' | \text{RSS}_\bp(k,b,t)\right)
\end{equation}

where $\mathds{P} \left(\bp = \bp' | \text{RSS}_{\bp'}(k,b,t)\right)$ represents the conditional probability of a user being in position $\bp'$ given the measurement in that position.
Expression \eqref{eq:2d_prob2} represents the intersection of multiple probabilities obtained from independently sampling the RSS distributions. Finally, to determine $\mathds{P} \left(\bp = \bp' | \text{RSS}_{\bp'}(k,b,t)\right)$, i.e., the conditional probability of a user standing in position $\bp'$ based on database information ($\widetilde{\text{RSS}}_{\bp'}(k,b)$) and a user measurement $\text{RSS}_{\bp'}(k,b,t)$. Based on our previous assumption that RSS values follow normal distributions, we can write

\vspace{-0.25cm}
\begin{align} \label{eq:2d_prob_assumption} 
    \mathds{P} \left(\bp = \bp' | \text{RSS}_\bp(k,b,t)\right) = \frac{1}{\sigma \sqrt{2 \pi} }\bigintss_{\text{RSS}_\bp(k,b,t) \hspace{-0.05cm} - \hspace{-0.05cm} \Delta}^{\text{RSS}_\bp(k,b,t) \hspace{-0.05cm} + \hspace{-0.05cm} \Delta} \hspace{-1.45cm} e^{-\frac{\left(x' - \widetilde{\text{RSS}}_\bp(k,b)\right)^2}{2 \sigma^2}} dx'
\end{align}
with $\sigma = \sigma_{def} = 2$ dBm and $\Delta << \mathsf{D}_{res}$ is half of the interval considered when accumulating probability and $DB_{res} = 10^{-3}$ dBm is the signal strength resolution in the DT database. This integral can be computed programmatically or by using the $erfc$ tables.

\section{Results and Discussion} \label{sec:Results}

\textbf{Baseline}: The baseline consists of a single user-measured RSS in the best-received beam in a given subband. To obtain the accuracy of this baseline, we sample 1000 times the RSS normal distribution $\text{RSS}_\bp(k,b,t) \sim \mathcal{N}\left(\mu_\bp(k,b), \sigma_\bp(k,b)\right)$ for all positions and crossed them with the respective RF map as described in Section \ref{sec:Solution}. The average localization error is shown in Figure \ref{fig:2D_acc_grid}. We see the lowest localization error in the LoS, especially where beams are stronger, and the highest error in NLOS, further away from the BS.

\textbf{Performance Predictability}:
The localization error and the received power (in the best beam and band) correlate at -0.87, suggesting a strong relation: higher received power leads to less error. This is evidence of the predictability of the approach, which is a desirable feature, especially given the difficulty of approximating NLOS performance bounds. Figure \ref{fig:2D_acc_grid} also shows positions 1 and 2. We use these positions as representatives of LoS and NLoS. Next, we compare the baseline localization accuracy in these positions with the accuracy achieved using more beams/subbands/times measured. 

\begin{figure}[t]
    \centering
    \includegraphics[width=.95\linewidth]{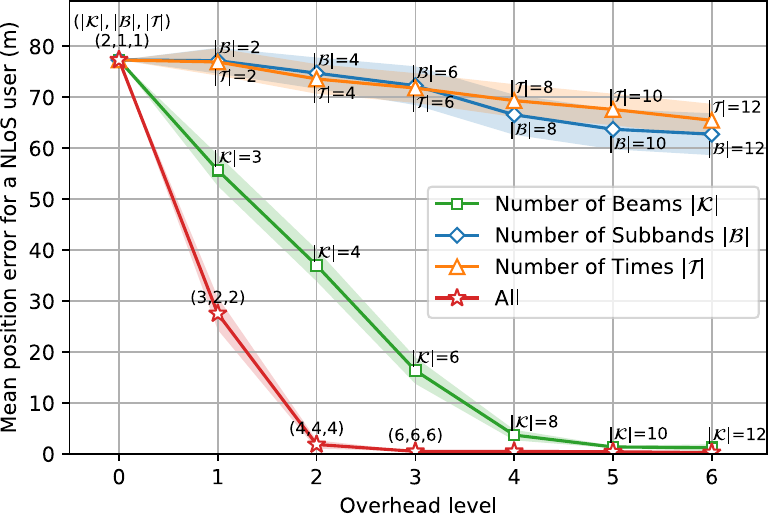}
    \caption{Impact of reporting parameter combinations on localization error for position 2 (NLoS). Localization error seems particularly affected by more RSS measurements in distinct beams. }
    \label{fig:pos2_multi}
    \vspace{-0.1cm}
\end{figure}

\textbf{Impact of More Measurements}: Figures \ref{fig:pos1_multi} and \ref{fig:pos2_multi} show the localization error depending on the number of beams $|\mathcal{K}|$, number of subbands $|\mathcal{B}|$, and number of times $|\mathcal{T}|$ measured and reported by the user. Each line shows the impact of each parameter individually (green, blue, and orange lines) and then jointly (red lines). As predicted, more measurements lead to more localization accuracy.

Comparing both figures, we see different impacts for each type of measurement. More samples across time are more useful in LoS than in NLoS. That effect can be attributed to higher RSS measurement noise, requiring more time averaging in NLoS due to less received power. The opposite happens with the number of beams: The performance improvement from reporting more beams is more significant in NLoS than in LoS. In LoS, the primary and secondary beams hold most location information, making subsequent beams less important. Since NLoS beams have more similar contributions, each additional beam reported provides more benefits. 

Overall, these results should be taken only as examples of the mean localization error in these two positions as the measurements increase. Since these RSS measurements are free to perform and practically free to report to the BS, obtaining sub-meter accuracies is promising. It should be noted, however, that more measurements may not always lead to better localization accuracy;  when the measurement noise exceeds the information gained by the measurement, considering more measurements may degrade positioning performance.

\begin{table}[t]
    \centering
    \label{tab:pos_acc}
    \caption{Maximum localization error (in meters) x\% of the time.}
    \begin{tabular}{cc|cccc}
    \multicolumn{2}{c|}{\multirow{2}{*}{\textbf{\begin{tabular}[c]{@{}c@{}}Percentage\\ of time\end{tabular}}}} & \multicolumn{4}{c}{\textbf{Parameter combinations (K,B,T)}} \\
    \multicolumn{2}{c|}{} & (1,1,1)       & (2,2,2)      & (4,4,4)      & (6,6,6)      \\ \hline \hline
    \multirow{3}{*}{\begin{tabular}[c]{@{}c@{}}Pos1\\ (LoS)\end{tabular}}                 & 99\%                & 48.1 m          & 31.7 m         & 2.1 m          & 1.8 m           \\
  & 90\%                & 40.8 m          & 9.4 m          & 2.0 m          & 1.6 m           \\
  & 80\%                & 32.3 m          & 3.0 m          & 1.9 m          & 1.6 m           \\ \hline
    \multirow{3}{*}{\begin{tabular}[c]{@{}c@{}}Pos2\\ (NLoS)\end{tabular}}                & 99\%                & 123.7 m        & 121.5 m        & 81.4 m         & 1.4 m          \\
  & 90\%                & 111.2 m        & 116.7 m          & 66.1 m          & 0.8 m           \\
  & 80\%                & 91.6 m         & 112.8 m          & 0.8 m          & 0.7 m           \\ \hline \hline
    \end{tabular}
\end{table}

\textbf{Localization Confidence}:
Figures \ref{fig:pos1_multi} and \ref{fig:pos2_multi} do not provide a localization confidence level; they only tell us the mean accuracy, which is insufficient to describe the distribution. We include in table \ref{tab:pos_acc} the cumulative distribution values of the horizontal localization error. Some noteworthy results include sub-meter accuracies in NLOS for 90\% of the time. Overall, considering the simulation uses a single access point, only communication signals, and at much larger distances than usual fingerprinting systems, still obtaining sub-meter accuracies motivates the scalability of this approach \cite{SurveyLocMethods, RSSI_to_CSI}.

\textbf{Limitations and Applicability}: The method presented in this paper is applicable to any wireless MIMO deployment using grid-of-beams or codebook-based beamforming - it needs to be adapted for codebook-free operation. It requires a 3D model of the environment (often available in, e.g., OpenStreetMaps), and its performance depends on the realism of the 3D model. Furthermore, shadowing is hard to consider in ray tracing simulations and must be accounted for via the parameter $\sigma$. For cases of extreme shadowing (e.g. 10-20 dB), the performance of the approach can be significantly affected, and further work is needed to cope with those cases. Moreover, the assessment lacked comparisons with other fingerprinting approaches. This is due to other approaches being used mainly indoors with smaller distances than those considered here, using several access points or not assessed with realistic ray tracing evaluations. Comparisons with real-world measurements are needed.

\textbf{Real-world Considerations}: Real measurements (using datasets such as DeepSense 6G \cite{DeepSense}) need to be considered in future studies to fully determine the potential of this approach. Some foreseeable challenges include excessive shadowing beyond the levels considered in our simulations. Also, real-world constraints can physically limit the number of reports. One way is receiver sensitivity limiting the number of detected beams. As such, overhead levels beyond level 3 may be unrealistic. Nonetheless, results show a promising trend in the performance and scalability of this approach, even for a small number of measurements.

\section{Conclusion} \label{sec:Conclusion}
This work proposed a novel way to save human effort while improving the performance of fingerprinting-based localization. We proposed digital twin RF maps, calculated with ray tracing in a digital replica of the target environment, to populate fingerprinting databases and localize users in the real world. The results showed meter-level accuracies 90\% of the time for NLOS users within 75m of the base station, even for relatively small databases using only communication signals. This work highlights the potential of digital twins for localization and the potential of fingerprinting strategies for massive outdoor user localization. The performance of this method in real-world scenarios remains to be investigated.

\balance

\end{document}